# Two-dimensional honeycomb borophene oxide: A promising anode material offering super high capacity for Li/Na-ion batteries

Junping Hu[a], Chengyong Zhong[b], Weikang Wu[c], Ning Liu[a], Yu Liu[a], Shengyuan A. Yang[c,e], Chuying Ouyang[d]*

a. School of Science, Nanchang Institute of Technology, Nanchang 330099, China

b. Institute for Advanced Study, Chengdu University, Chengdu 610106, China

c. Research Laboratory for Quantum Materials, Singapore University of Technology and Design, Singapore 487372, Singapore

d. Department of Physics, Laboratory of Computational Materials Physics, Jiangxi Normal University, Nanchang 330022, China.

e. Center for Quantum Transport and Thermal Energy Science, School of Physics and Technology, Nanjing Normal University, Nanjing 210023, China

**ABSTRACT:** Rational design of novel two-dimensional (2D) electrode materials with high capacity is crucial for the further development of Li-ion and and Na-ion batteries. Herein, based on first-principles calculations, we systemically investigate Li and Na storage behaviors in the recently discovered 2D topological nodal-loop metal — the honeycomb borophene oxide ($h$-$B_2O$). We show that $h$-$B_2O$ is an almost ideal anode material. It has good conductivity before and after Li/Na adsorption, fast ion diffusion with diffusion barrier less than 0.5 eV, low open-circuit voltage (< 1 V), and small lattice change (<6.2%) during intercalation. Most remarkably, its theoretical storage capacity is extremely high, reaching up to 2137 mAh·$g^{-1}$ for Li and 1425 mAh·$g^{-1}$ for Na. Its Li storage capacity is more than six times higher than graphite (~372 mAh·$g^{-1}$), and is actually the highest among all 2D materials discovered to date. Our results strongly suggest that 2D $h$-$B_2O$ is an exceedingly promising anode material for both Li- and Na-ion batteries with super high capacity.

# 1. Introduction

Energy storage is one of the most rapidly developing industries today. There is huge demand for energy storage technology in portable electronic devices, electric vehicles, and large-scale energy storage power stations, where rechargeable metal-ion batteries, especially the Li-ion batteries (LIBs) and Na-ion batteries (NIBs), have found wide applications.[1-5] Constant efforts have been devoted to improve the performance of rechargeable batteries, including the aspects of reversible capacity, power density, and long cycle life.[6-9] Among these, probably the most important challenge at present is on the capacity, which severely limits the miniaturization of the devices. To tackle this challenge, one promising approach is to develop new electrode materials with high capacity to replace the existing ones.

The rise of two-dimensional (2D) materials offers new possibilities for this line of research. With only atomic thickness, 2D materials enjoy the advantage of extremely high surface-to-volume ratio, which is beneficial for enhancing the capacity. Besides, the metal ion diffusion on the surface of 2D materials is generally fast, and 2D materials usually have relatively small volume change during battery charging and discharging. [10] All these are desired features for a good electrode material. Up to now, many 2D materials have been proposed as candidate electrode materials for LIBs and NIBs, including: (i) graphene and related materials;[11-13] (ii) transition-metal dichalcogenides, such as $MoS_2$,[14] $WS_2$,[15] and $VS_2$;[16] (iii) transition-metal carbides (also known as MXenes), such as $Mo_2C$,[17] $Ti_3C_2$,[18] and $V_2C$;[19] (iv) borophene[20], phosphorene,[21] and related systems;[22-24] (v) other 2D systems, such as metal nitrides,[25] metal oxides[26] and so on.[27-30] Among these 2D materials, borophene is found to possess an extremely high capacity due to its small atomic mass.[31] We note that graphene itself has negligible capacity because its interaction with Li or Na is too weak. To improve its capacity, modification methods such as silicon or germanium doping have been proposed, which indeed promote the interaction between metal ions and the host and achieve higher storage capacity[10, 32]. Since boron is just neighboring to carbon in the periodic table, it is likely that similar modification methods applied to borophene may further enhance its capacity.

In general, 2D materials are susceptible to oxidation under ambient conditions. In many cases, this is considered as a disadvantage. However, from another point of view, oxidation can help to improve the stability of the structure and to customize physical and chemical properties.[33] Recently, Li et al. successfully synthesized monolayer honeycomb borophene in experiment. Zhong et al. found that upon oxidation, the material would transform into a new 2D material, the honeycomb borophene oxide ($h$-$B_2O$), which is predicted to be stable up to 1000 K[34]. Interestingly, 2D $h$-$B_2O$ is found to be a 2D topological nodal-loop metal, which is expected to have high conductivity due to its low-energy topological fermions with linear dispersion. Further considering its low mass density, one may naturally wonder: Can 2D $h$-$B_2O$ be a superior electrode material for LIBs or NIBs?

Motivated by the previous discussions, here, based on first-principles calculations, we investigate Li and Na storage properties in 2D $h$-$B_2O$. We study the Li/Na adsorption and migration processes on 2D $h$-$B_2O$, and evaluate the storage capacities and the corresponding average open-circuit voltages. Our result shows that 2D $h$-$B_2O$ is an almost ideal anode material for both LIBs and NIBs, with good conductivity before and after Li/Na adsorption, low diffusion barrier, low average open-circuit voltage, and small lattice change. Most remarkably, its Li and Na storage capacity can reach up to 2137 mAh·g$^{-1}$ and 1425 mAh·g$^{-1}$, respectively. The value for Li capacity is six times higher than graphite (~372 mAh·g$^{-1}$) and is in fact the highest among all 2D materials studied to date, to the best of our knowledge. Our result thus suggests the great potential of 2D $h$-B$_2$O as an anode material for LIBs and NIBs with super high capacities.

## 2. Computational methods

Our first-principles calculations are based on the density functional theory (DFT) using the plane-wave pseudopotentials[35-36] as implemented in the Vienna Ab initio Simulation Package (VASP) [37-38]. The exchange-correlation functional is modeled in the generalized gradient approximation (GGA) with the Perdew-Burke-Ernzerhof (PBE)[39-40] realization. Carbon $2s^22p^2$ electrons are treated as valence electrons in all the calculations. A cutoff energy of 520 eV is employed for the plane wave expansion of the wave functions. The Brillouin zone is sampled with 3 × 3 × 1 Monkhorst-Pack $k$-point mesh[41] for the structural optimization, and with 5 × 5 × 1 mesh for the electronic structure calculations. The convergence criteria for the total energy and ionic forces are set to be 10$^{-5}$ eV and 10$^{-3}$ eV/Å, respectively.

## 3. Results and discussion

### 3.1 Lattice structure

The lattice structure for 2D $h$-B$_2$O is schematically shown in Fig. 1(a) and 1(b). For the 3 × 3 supercell containing 18 B and 9 O atoms (which will be used for the metal-ion adsorption calculation below), the optimized lattice parameter are a=b=11.8095 Å with a rhombic cell angle of 139.001°. The equilibrium bond lengths for B-O and B-B bonds are 1.343 Å and 1.704 Å, respectively, which are consistent with the previous work[34]. As shown in Fig. 1(a), the atomic structure of 2D $h$-B$_2$O is completely plat with single-atom thickness. Its space group is *Cmmm* (No. 65). The charge density distribution and electron localization function (ELF) are plotted in Fig. 1(c) and 1(d). One can observe that most of the charge is distributed around the oxygen atom, and the strong bonds are formed between B and O and between B and B.

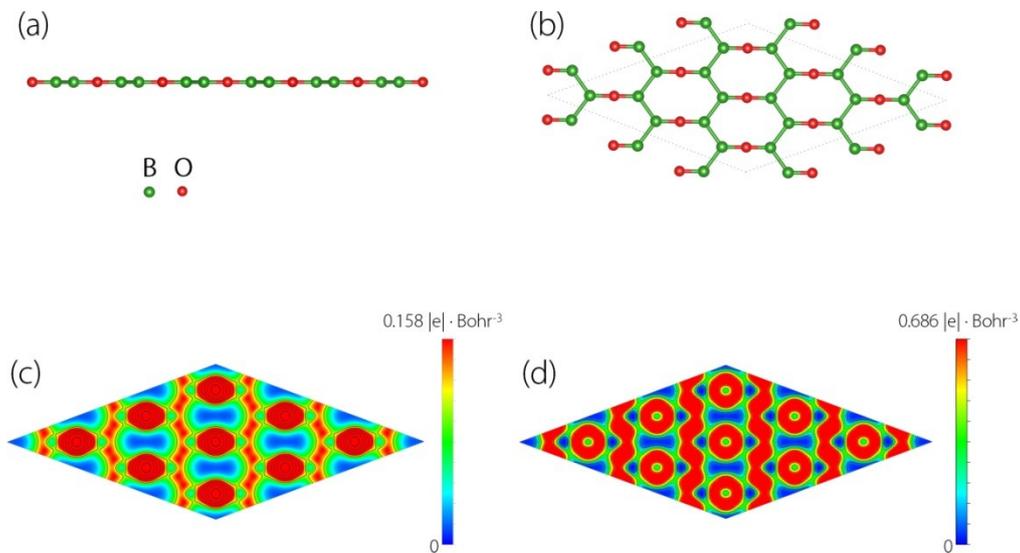

Figure 1: (a) and (b) are the side and top views of the optimized structure for 2D honeycomb borophene oxide ($h$-B$_2$O). (c) shows the charge density distribution, and (d) shows the electron localization function (ELF). In (a) and (b), the green balls and red balls represent the boron and oxygen atoms, respectively, and the grey dashed lines in (b) indicate the 3 × 3 supercell.

### 3.2 Lithium and sodium adsorption on $h$-B$_2$O

To check whether 2D $h$-B$_2$O is suitable to be cathode or anode material for Li/Na-ion batteries, it is necessary to investigate the adsorption behavior of a single Li or Na atom on the material. As shown in Fig. 2(e), three kinds of adsorption sites are considered: i) hollow site, marked as C1, which is above the center of the B-O ring; ii) top sites, marked as T1 and T2, which are respectively above the B and O atoms; iii) bridge sites, marked as B1 and B2, which are respectively above the mid-points of the B-O bond and the B-B bond.

The adsorption behavior is characterized by the adsorption energy, defined as

$$E_{ad} = E_{B_{18}O_9} - E_{B_{18}O_9 M} - E_M, \qquad (1)$$

where $M$ represents Li or Na, $E_M$ is the cohesive energy of the bulk Li or Na metal, $E_{B_{18}O_9}$ and $E_{B_{18}O_9 M}$ are the total energies of $h$-$B_2O$ (of a 3 × 3 supercell) before and after $M$ adsorption. All the total energies are obtained after performing fully structural optimizations.

A negative value of $E_{ad}$ indicates that the metal atom prefers to adsorb on 2D $h$-$B_2O$ instead of forming a metal cluster. Conversely, a positive value indicates the tendency for forming a metal cluster directly, without adsorption. According to our calculation results, we identify the most stable configurations for Li and Na adsorption, as plotted in Fig 2(c) and 2(d), respectively. We find that both atoms prefer the hollow site above the center of the B-O ring, with a distance of 1.35 Å and 1.92 Å from the 2D plane for Li and Na, respectively. The corresponding adsorption energies are -0.684 eV for Li and -0.477 eV for Na. These values (in absolute value) are less than 1.5 eV, implying that 2D $h$-$B_2O$ is suitable as an anode material for Li and Na ion batteries. By carrying out the Bader charge analysis, as shown in Table 1, we find that approximately 0.87 e and 0.84 e are transferred to the $B_{18}O_9$ supercell after Li and Na adsorption, respectively. To better understand the interaction between Li/Na and the host, we also plot the charge density difference profiles in Fig. 2(f) and 2(g). One observes that electrons in both cases tend to accumulate around the B-O ring, and their density surrounding the Li or Na atom decreases. All these results indicate a strong interaction between Li/Na and 2D $h$-$B_2O$, which is beneficial for obtaining a high capacity.

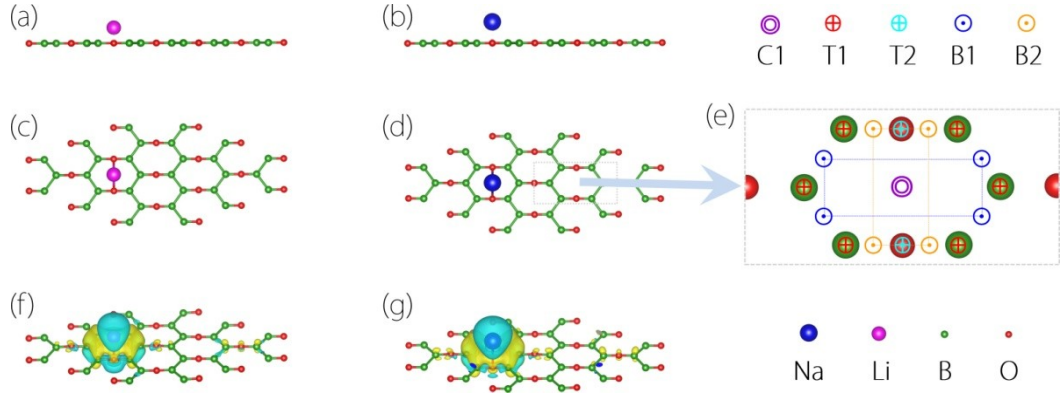

Figure 2: (a) and (b) show the side views of the most stable Li and Na adsorption configurations, respectively. (c) and (d) are the corresponding top views. (f) and (g) show the corresponding charge density difference profiles. The blue and the yellow colors represent regions with electron depletion and accumulation, respectively. (e) shows the considered Li/Na adsorption sites on the surface of 2D $h$-$B_2O$ (top view).

Next, we execute the electronic structure calculations to check the metallicity of 2D $h$-$B_2O$ during the above adsorption process. As a battery electrode material, it is desired that the material remains metallic before and after adsorption. For 2D $h$-$B_2O$, it has been revealed in Ref.… that the material (without adsorption) is a topological nodal-loop metal with high Fermi velocity. The metallicity is confirmed here in Fig. 3(a), which plots the density of states (DOS) for $B_{18}O_9$ (i.e. $h$-$B_2O$ with a 3 × 3 supercell). Similarly, we calculate the DOS for $h$-$B_2O$ with Li/Na adsorption (corresponding to $B_{18}O_9Li$/$B_{18}O_9Na$), and the results are shown in Fig. 3(b) and 3(c). One finds that the system remains metallic after adsorption, with a considerable DOS at the Fermi level. The metallic properties of 2D $h$-$B_2O$ and its Li/Na embedded states ensure good electrical conductivity, paving the way for its application as an anode material.

Table 1: Bader charge analysis

|  | Average Charge State | | | |
|---|---|---|---|---|
|  | B | O | Li | Na |
| $B_{18}O_9$ | 2.243 | 7.514 | | |
| $LiB_{18}O_9$ | 2.287 | 7.522 | 0.133 | |
| $Li_{18}B_{18}O_9$ | 2.990 | 7.671 | 0.174 | |
| $NaB_{18}O_9$ | 2.285 | 7.524 | | 0.160 |
| $Na_{18}B_{18}O_9$ | 2.825 | 7.585 | | 0.382 |

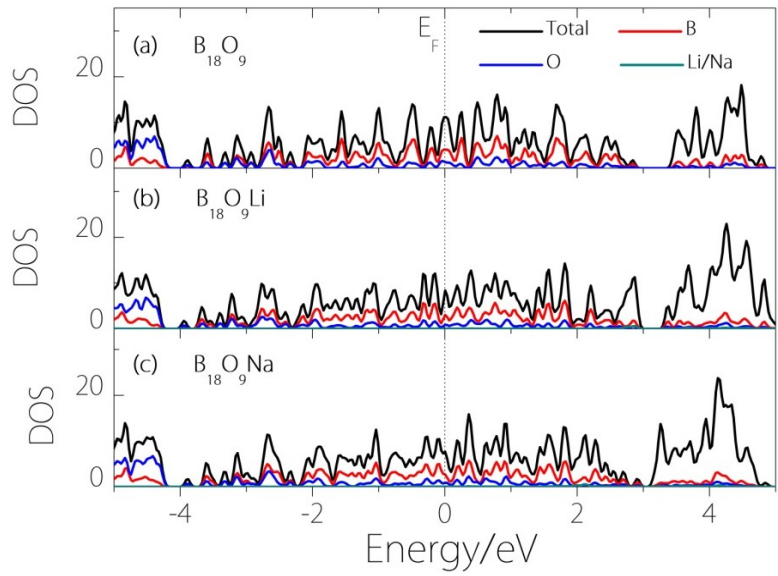

Figure 3: Density of states (DOS) for 2D $h$-$B_2O$ before and after Li/Na adsorption: (a) $B_{18}O_9$, (b) $B_{18}O_9Li$, and (c) $B_{18}O_9Na$.

### 3.3 Diffusion of Li/Na on the surface of $h$-$B_2O$

Rate performance is another key characteristic of rechargeable batteries. To have a good rate performance, the energy barrier for Li or Na diffusion must be low. Here, we carry out the diffusion barrier calculations, namely, we examine and optimize the diffusion paths and compute their corresponding diffusion barriers by using the climbing-image nudged elastic band (CL-NEB) method.[42] For the 2D $h$-$B_2O$, Based on the results obtained in the last section, we consider the four pathways as shown in Fig. 4(a). Here, the initial and the final states correspond to the most energetically stable Li/Na adsorption configurations found in the last section. The corresponding energy barriers for Li and Na migration are plotted in Fig. 4(b) and 4(c), respectively. The the saddle points for each path are identified. One finds that for both cases, the Path2 possesses the lowest diffusion barrier, which is 0.45 eV for Li and 0.17 eV for Na. These barrier values are lower than that for graphite (~0.6 eV), hence a fast ion diffusion and a good rate performance can be expected for 2D $h$-$B_2O$.

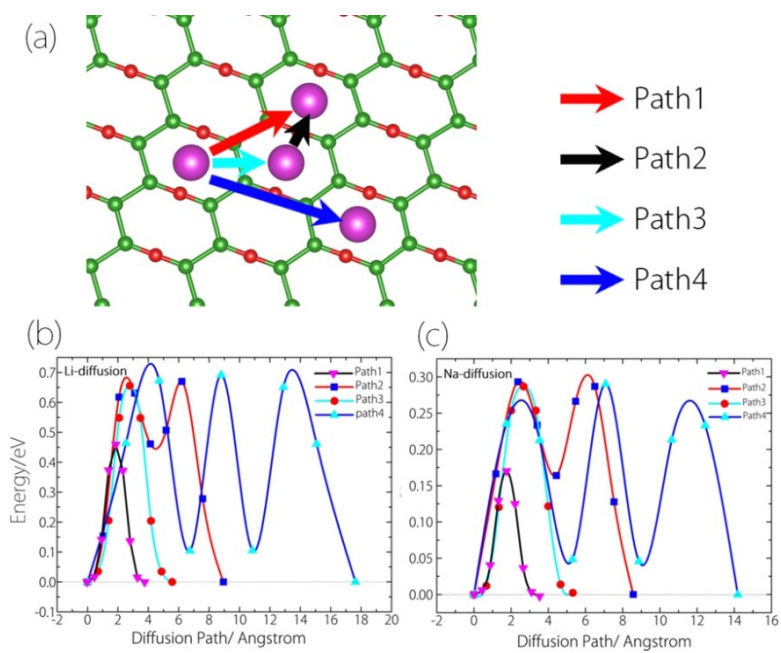

Figure 4: (a) The considered Li/Na migration pathways on 2D $h$-$B_2O$. Note that after optimization, each path may deviate from the simple straight line illustrated in the figure. (b) The diffusion barrier profiles for Li migration for the considered four paths. (c) The corresponding barrier profiles for Na migration.

## 3.4 Theoretical capacity and average open-circuit voltage

As we have discussed, the storage capacity is a key indicators for the performance of an electrode material. To evaluate the capacity, we now increase the concentration of the Li/Na adsorption on the 2D $h$-B$_2$O monolayer. In the calculation, we still use the 3 × 3 supercell, and Li or Na atoms are intercalated on both sides of the host one by one. We here simulate the following half-cell reaction vs $M/M^+$ ($M$ = Li, Na):

$$B_{18}O_9 + xM^+ + xe^- \leftrightarrow B_{18}O_9M_x \tag{2}$$

where $x$ is the number of the intercalated atoms.

The average open-circuit voltage can be obtained by computing the total energies before and after $M$ intercalation. To obtain more accurate energy values, both the lattice parameters and atomic positions are full relaxed for the intercalated configurations. Usually the entropy and volume effects are negligible during the reaction. It follows that the average open-circuit voltage for the intercalation reaction involving $xM^+$ ions can be calculated from the total energies difference as following:

$$V_{ocv} = (E_{B_{18}O_9} + xE_M - E_{B_{18}O_9M_x})/x \tag{3}$$

where $E_M$ is the cohesive energy in the bulk Li or Na metal; $E_{B_{18}O_9}$ and $E_{B_{18}O_9M_x}$ are the total energies of 2D $h$-B2O before and after $M$ intercalation.

As mentioned before, according to the adsorption energies, 2D $h$-B$_2$O is suitable as anode material for both LIBs and NIBs. We have found that the position of adsorbed Li/Na atoms in the first layer are at the hollow sites above the centers of B-O rings (see Fig. 5(a)). The Li/Na can be symmetrically adsorbed on both sides of the $h$-B$_2$O monolayer, as shown in Fig. 5. After the first layer adsorption, via calculation, we find that the second layer Li/Na atoms can be adsorbed on the top sites above the B atoms (see Figure 5…). By adding the adsorbed atoms one by one, we calculate the sequential adsorption energy, defined as

$$E_{sae} = E_{B_{18}O_9M_{n+1}} - E_{B_{18}O_9M_n} - E_M, \tag{4}$$

where $E_{B_{18}O_9M_{n+1}}$ and $E_{B_{18}O_9M_n}$ are the total energies of the $h$-B$_2$O monolayer with n+1 and n Li or Na adsorbed atoms. In the calculation, the negative value of $E_{sae}$ would indicate that the corresponding Li/Na atom can still be intercalated. The calculate proceeds until $E_{sae}$ becomes positive, meaning that no more Li/Na can be accommodated and the metal atoms would form clusters. Our calculation shows that the 3 × 3 supercell of $h$-B$_2$O could accommodate up to 27 Li atoms, which amounts to one and half adsorption layers as shown in Fig.… It corresponds to the chemical stoichiometry of B$_{18}$O$_9$Li$_{27}$. The evaluated average open-circuit voltage decreases from 0.68 V to 0.34 V with the increase of the adsorbed Li layers. Then, we can obtain the theoretical capacity (*C*) through the following equation,

$$C_{theo} = xF / 3.6 M_{h-B_2O}, \tag{5}$$

where *x* is the maximum number of electrons involving the half-cell reaction, *F* is the Faraday constant, and $M_{h-B_2O}$ is the mass of $h$-B$_2$O in units of g·mol$^{-1}$. The calculated theoretical capacity for LIBs is 2137 mAh·g$^{-1}$, which is extremely high.

Following the similar method, for NIBs, we find that the $h$-B$_2$O monolayer can at most host one year of Na atoms, namely, its 3 × 3 supercell can accommodate up to 18 Na atoms, which corresponds to a chemical stoichiometry of $B_{18}O_9Na_{18}$. The evaluated average open-circuit voltage decreases from 0.48 V to 0.12 V with the increase of the adsorbed Na layers. The corresponding theoretical capacity for NIBs is 1425 mAh·g$^{-1}$.

Furthermore, we examine the ELFs for the Li or Na adsorption configurations, as shown in Fig. 5(e) and 5(f). One observes that although the concentrations of electrons forming negatively charged electron cloud vary in space, the electrons distributed among Li or Na atoms contribute a net negatively charged environment, which helps to generate an attractive interaction between the Li or Na atomic layer and the host material, stabilizing the adsorption configurations. This observation also indicates that the above results are reasonable.

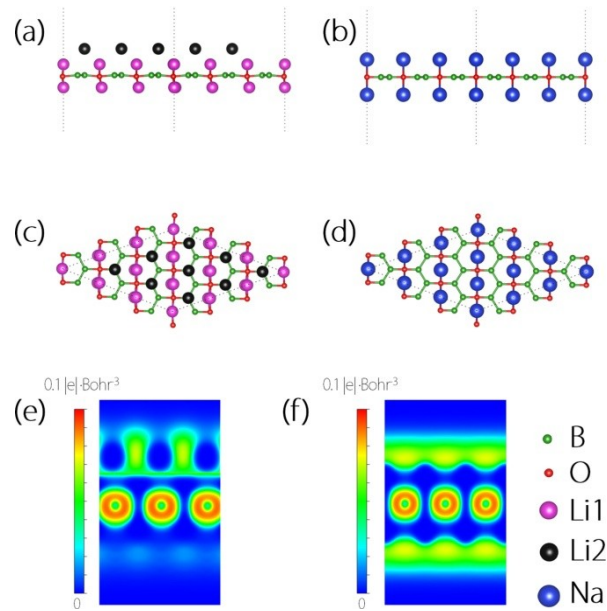

Figure 5: Structures with maximal Li or Na intercalation. (a) and (c) are the side and top views of $B_{18}O_9Li_{27}$. Here, the black and pink balls denote Li atoms in different layers. (b) and (d) are the side and top views of $B_{18}O_9Na_{18}$. (e) and (f) are the corresponding electron localization functions (ELFs).

In addition, to evaluate the cycling performance, we carry out the calculations on the change of the lattice constants. Our results show that, during the intercalation process, the lattice constants in the $x$-$y$ plane only experience a tensile strain about 6.2% for Li and 4.1% for Na, which are considered to be small (typically values below 10% are acceptable). This offers another advantage of $h$-$B_2O$ to be utilized as anode material for both LIBs and NIBs.

**3.5 Comparison with other 2D anode materials on capacity**

As we have shown, 2D $h$-$B_2O$ has super high capacity for Li and Na storage, which can reach 2137 mAh·g$^{-1}$ for Li and 1425 mAh·g$^{-1}$ for Na. Here, we make a comparison of these values to those for other 2D materials studied in previous works.[43-70] The results are plotted in Fig. 6. Remarkably, one observes that the Li storage capacity of 2D $h$-$B_2O$ is the highest among all 2D materials studied to date. Regarding Na capacity, the value is only lower than that for the $\beta_{12}$-borophene (~1984 mAh·g$^{-1}$)[31] but is even larger than the capacity for $\chi_3$-borophene (~1240 mAh·g$^{-1}$) [31]. The super high storage capacity of $h$-$B_2O$ can be attributed to two factors: (1) the large periodic voids in the flat lattice plane, which increases the adsorption stability and lowers the mass density; and (2) the small atomic mass of boron and oxygen. In addition, the mechanism of multilayer adsorption, which is a key advantage for many 2D materials, further boosts the Li storage capacity of $h$-$B_2O$.

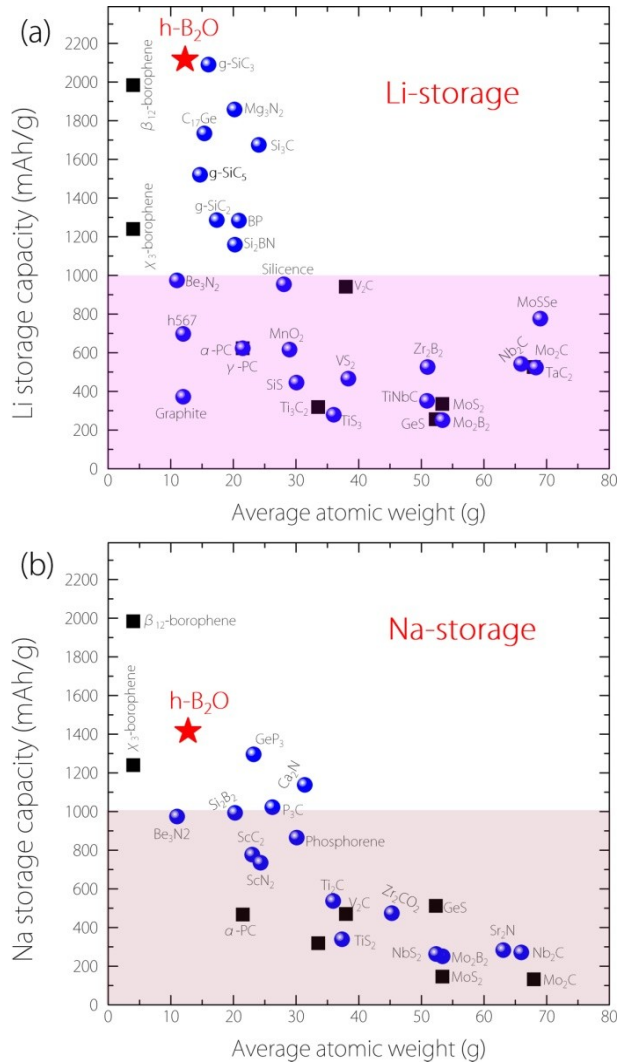

Figure 6: Comparison of the theoretical Li and Na storage capacity values between *h*-B$_2$O and other 2D anode materials studied in literature (plus graphite).

## 4. Conclusion

In this work, we discover that the 2D honeycomb borophene oxide (*h*-B$_2$O) is a very promising anode material for both LIBs and NIBs with super high capacity. Our results show that 2D *h*-B$_2$O has good conductivity before and after Li/Na adsorption, low Li/Na diffusion barrier (< 0.5 eV), low average open-circuit voltage, and small lattice change (< 6.2%). In particular, we find the Li and Na theoretical storage capacities can reach up to 2137 mAh·g$^{-1}$ and 1425 mAh·g$^{-1}$, respectively. The Li capacity is six times higher than graphite (372 mAh·g$^{-1}$) and is the highest among all 2D materials studied to date. Our work reveals *h*-B$_2$O as an exceedingly promising anode material for both LIBs and NIBs, and further provides insights on engineering 2D materials, especially 2D topological materials, to achieve enhanced capacities.

**Acknowledgments**

The authors thank D. L. Deng for valuable discussions. This work is supported by the Natural Science Foundation of China (NSFC Grant No. 11564016), Natural Science Foundation of Jiangxi Province (Grants No. 20171BAB216014), Scientific Research Fund of Jiangxi Provincial Education Department (GJJ171010), and Singapore Ministry of Education Academic Research Fund Tier 2 (MOE2017-T2-2-108).

405-411.

## Table of Contents Graphic

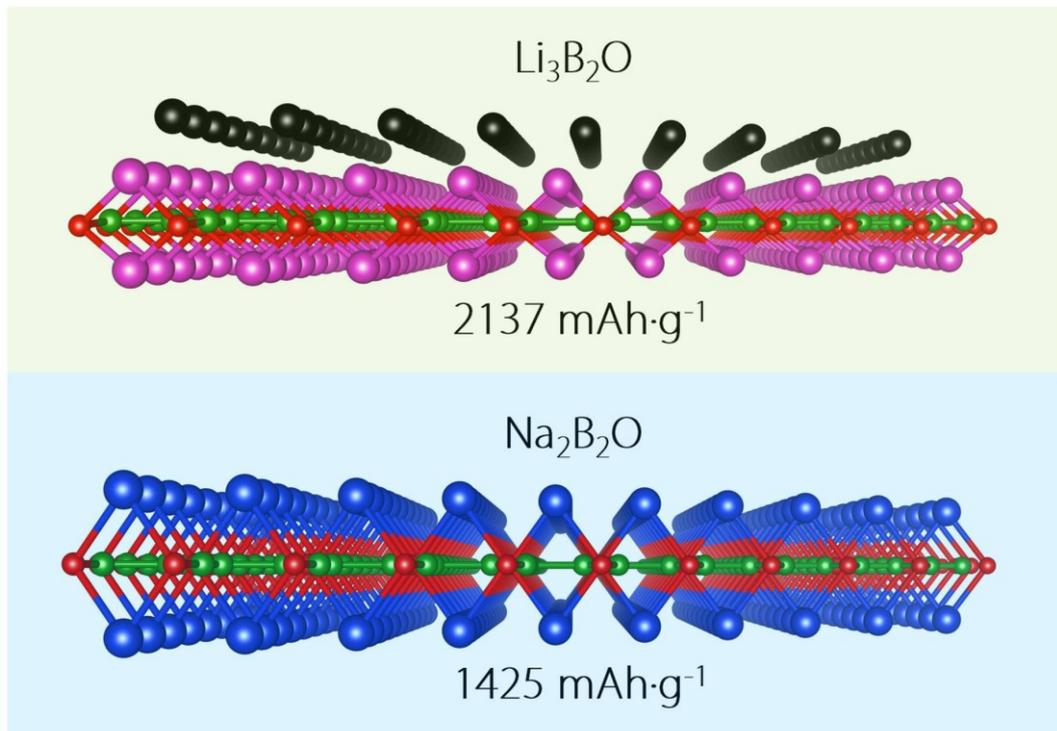